\documentclass[aps,pre,twocolumn,amsfonts,amssymb,superscriptaddress,showpacs]{revtex4-1}
\usepackage{placeins}
\usepackage{graphicx}
\usepackage{amsmath}
\usepackage{times}
\usepackage{color}
 \usepackage{hyperref}

\newcommand{\eq}[1]{Eq.~(\ref{eq:#1})}

\newcommand{\fig}[1]{Fig.~\ref{fig:#1}}

\begin{document}
\mbox{}\\[-15mm]
\hspace*{0.2\textwidth}
{\href{http://pre.aps.org}{Physical Review E}, in print (2012)}
\\[-50mm]
\title{Stochastic differential equations for evolutionary dynamics with demographic noise and mutations}
\author{Arne Traulsen}
\affiliation{Evolutionary Theory Group, Max-Planck-Institute for Evolutionary Biology, August-Thienemann-Str. 2, 24306 Pl{\"o}n, Germany}
\author{Jens Christian Claussen}
\affiliation{Institute for Neuro- and Bioinformatics, University of L{\"u}beck, 23538 L{\"u}beck, Germany}
\author{Christoph Hauert}
\affiliation{Department of Mathematics, The University of British Columbia, 1984 Mathematics Road, Vancouver, B.C., Canada, V6T 1Z2}
\date{March 12, 2012}
\begin{abstract}
We present a general framework to describe the evolutionary dynamics of an arbitrary number of types in finite populations based on stochastic differential equations (SDE). 
For large, but finite populations this allows to include demographic noise without requiring explicit simulations.
Instead, the population size only rescales the amplitude of the noise.
Moreover, this framework admits the inclusion of mutations between different
types, provided that mutation rates, $\mu$, are not too small compared to the 
inverse population size $1/N$. This ensures that all types are almost always represented in the population and that the occasional extinction of one type does not result in an extended absence of that type. 
For $\mu N\ll1$ this limits the use of SDE's, but in this case there are well established alternative approximations based on time scale separation.
We illustrate our approach by a Rock-Scissors-Paper game with mutations,
where we demonstrate excellent agreement with simulation based results for sufficiently large populations.
In the absence of mutations the excellent agreement extends to small population sizes.
\end{abstract}
\pacs{
87.23.-n, 		
89.65.-s 		  
02.50.Ey 		
}
\maketitle

\section{Introduction}
Populations evolve when different individuals have different traits or strategies that determine their reproductive success.
In population genetic models, this success is typically constant, whereas in evolutionary game theory the fitness of an individual depends on interactions with other members of the population.
Consequently,
fitness depends on the relative proportions (or frequencies)
of different strategic types and hence gives rise to 
so called frequency dependent selection.
More successful strategies increase in abundance and may either take over the entire population or, as the strategy becomes increasingly common, 
may suffer from a decrease in fitness, which can result in the co-existence of two or more traits within the population. For example, in host-parasite coevolution, a rare parasite may be most successful, but once it reaches high abundance, selection pressure on the host increases and the host is likely to develop some defense mechanism. 
Consequently, the success of the parasite decreases. 
Such dynamics can be conveniently described by evolutionary game theory \cite{maynard-smith:1982to,hofbauer:1998mm,nowak:2006bo,szabo:2007aa,roca:2009aa,perc:2010bb}.   

Traditionally, the mathematical description of evolutionary game dynamics is 
formulated in terms of the deterministic replicator equation \cite{hofbauer:1998mm}. This implies that population sizes are infinite and populations are unstructured. 
Only more recently the stochastic dynamics in finite populations attracted increasing attention \cite{nowak:2004pw,traulsen:2005hp}. 
Typically, the stochastic approach becomes deterministic in the limit of infinite populations. 
For large, but finite populations the dynamics can be approximated by stochastic differential equations (SDE) \cite{helbing:1996aa,traulsen:2005hp,traulsen:2006hp}. 
This analytic approach is a natural extension of the replicator dynamics, which is capable of bridging the gap between deterministic models and individual based simulations. 
Moreover, SDE's are typically computationally far less expensive than simulations because the execution time does not scale with population size. 
\\[-8mm]

\section{Master equation}
In unstructured,
finite populations of constant size, $N$, consisting of $d$ distinct strategic types and with a mutation rate, $\mu$, evolutionary changes can be described by the following class of birth-death processes: 
In each time step, one individual of type $j$ produces a single offspring and displaces another randomly selected individual of type $k$. 
With probability $1-\mu$, no mutation occurs and $j$ produces an offspring of the same type. 
But with probability $\mu$, the offspring of an individual of type $i$ ($i\neq j$) mutates into a type $j$ individual. This results in two distinct ways to increase the number of $j$ types by one at the expense of decreasing the number of $k$ types by one, hence keeping the population size constant. Biologically, keeping $N$ constant implies that the population has reached a stable ecological equilibrium and assumes that this equilibrium remains unaffected by trait frequencies. 
The probability for the event of replacing a type $k$ individual with a type $j$ individual is denoted by $T_{kj}$ and is a function of the state of the population $\boldsymbol X=(X_1, X_2, \ldots X_d)$, with $X_n$ indicating the number of individuals of type $n$ 
such that $\sum_{n=1}^d X_n = N$.

For this process it is straight forward to write down a Master equation \cite{traulsen:2006hp} and, at this point, there is no need to further specify the transition probabilities $T_{kj}$.
\\[-5mm]
\begin{align}
P^{\tau+1}\!({\boldsymbol X})\! =\ \!& P^{\tau}\!({\boldsymbol X})\! +\!\!\! \sum_{j,k=1}^d\!\! P^{\tau}({\boldsymbol X}_j^k)  T_{kj}({\boldsymbol X}_j^k)\! -\! P^{\tau}({\boldsymbol X})T_{jk}({\boldsymbol X}))
\label{master}
\end{align}
\mbox{}\\[-8mm]
where $P^\tau({\boldsymbol X})$ denotes the probability of being in state $\boldsymbol X$ at time $\tau$ and ${\boldsymbol X}_j^k=(X_1, \ldots X_j-1, \ldots X_k+1, \ldots X_d)$ represents a state adjacent to $\boldsymbol X$.

\section{Fokker-Planck and Langevin equations}
While the Master equation~(\ref{master}) is rather unwieldy, a Kramers-Moyal expansion yields a convenient approximation for large but finite $N$ in the form of a Fokker-Planck equation \cite{gardiner:2004aa}
\begin{align}
\dot \rho({\boldsymbol x})  = 
- \sum_{k=1}^{d-1} \frac{\partial}{\partial x_k} \rho({\boldsymbol x}) {\cal A}_{k}({\boldsymbol x}) 
+ \frac{1}{2}\sum_{j,k=1}^{d-1} \frac{\partial^2}{\partial x_k \partial x_j}\rho({\boldsymbol x}) 
{\cal B}_{jk}({\boldsymbol x})
\label{eq:FPE}
\end{align}
where ${\boldsymbol x} = {\boldsymbol X}/N$ represents the state of the population in terms of frequencies of the different strategic types and $\rho({\boldsymbol x})$ is the probability density in state ${\boldsymbol x}$. 
Due to the normalization $\sum_{k=1}^d x_k =1$, 
it suffices to consider $d-1$ elements of the deterministic drift vector ${\cal A}_{k}({\boldsymbol x})$, $k=1,\ldots d-1$. Similarly, we only need to consider a diffusion matrix ${\cal B}_{jk}({\boldsymbol x})$ with dimension $(d-1)\times(d-1)$, i.e.\ $j,k=1,\ldots d-1$. 
The drift vector ${\cal A}_{k}({\boldsymbol x})$ is given by
\begin{equation}
{\cal A}_{k}({\boldsymbol x}) = \sum_{j=1}^{d} \Big(T_{j k}({\boldsymbol x}) - T_{k j}({\boldsymbol x}) \Big) = -1+\sum_{j=1}^{d} T_{j k}({\boldsymbol x}).
\label{eq:driftvec}
\end{equation}
For the second equality we have used $\sum_{j=1}^d T_{kj}({\boldsymbol x})=1$, which simply states that a $k$-type individual transitions to some other type (including staying type $k$) with probability one. 
${\cal A}_{k}({\boldsymbol x})$ is bounded in $[-1, d-1]$ because the $T_{jk}$ are probabilities. 

The diffusion matrix ${\cal B}_{jk}({\boldsymbol x})$ is defined as
\begin{subequations}
\begin{eqnarray}
{\cal B}_{jk}({\boldsymbol x}) 
&=& - \frac{1}{N} \left[  T_{j k}({\boldsymbol x}) + T_{k j}({\boldsymbol x}) \right] \quad {\rm for} \quad j \neq k  \\
{\cal B}_{jj}({\boldsymbol x}) 
&=&   \frac{1}{N} \left[  \sum_{l=1,l\neq j}^d
\Big(
T_{j l}({\boldsymbol x})+T_{l j}({\boldsymbol x}) \Big) \right].
\end{eqnarray}%
\label{eq:diffmat}%
\end{subequations}
For a detailed derivation see e.g.\ \cite{traulsen:2006hp} or \cite{bladon:2010ws}. 
Note that the diffusion matrix is symmetric, 
${\cal B}_{jk}({\boldsymbol x}) = {\cal B}_{kj}({\boldsymbol x})$ and vanishes as $\sim 1/N$ in the limit $N\to\infty$.
Moreover, we have for all diagonal elements ${\cal B}_{jj}({\boldsymbol x})\geq 0 $
and for the non-diagonal elements ${\cal B}_{jk}({\boldsymbol x}) \leq 0$. 
Note that $T_{jj}({\boldsymbol x})$ cancels in \eq{driftvec} and does not appear in \eq{diffmat}
-- it is thus of no further concern.

The noise of our underlying process
is uncorrelated in time and hence the It\^o calculus \cite{gardiner:2004aa} can be applied to derive a Langevin equation, which represents in our case a stochastic replicator-mutator equation,
\begin{equation}
\dot x_k = {\cal A}_k({\boldsymbol x}) + \sum_{j=1}^{d-1} {{\cal C}_{kj}({\boldsymbol x})} \xi_j(t) 
\label{eq:langevin}
\end{equation}
where the $\xi_j(t)$ represent uncorrelated Gaussian white noise with unit variance, $\langle \xi_k(t) \xi_j(t^\prime) \rangle = \delta_{kj} \delta(t-t^\prime)$. 
The matrix ${{\cal C}({\boldsymbol x})}$ is defined by ${\cal C}^T({\boldsymbol x}) {\cal C}({\boldsymbol x}) =  {\cal B}({\boldsymbol x})$ and its off-diagonal elements are responsible for correlations in the noise of different strategic types. 

Fluctuations arising in finite populations are approximated by the 
stochastic term in the Langevin equation (\ref{eq:langevin}).
For given transition probabilities the matrix ${\cal C}({\boldsymbol x})$ provides a quantitative description of the fluctuations introduced by microscopic processes in finite populations. In the limit $N\to\infty$ the matrix ${\cal C}({\boldsymbol x})$ vanishes with $\sim 1/\sqrt{N}$ and we recover a deterministic replicator mutator equation. 
Note that the replicator equation does not impose an upper or lower bound on ${\cal A}$ (c.f. \eq{driftvec}). 
However, this difference merely amounts to a (constant) rescaling of time.

The multiplicative character of the noise and its strategy-strategy correlations are determined by the form of the matrix ${{\cal C}{}}({\boldsymbol x})$. 
In order to determine ${{\cal C}{}}({\boldsymbol x})$, we first diagonalize ${\cal B}({\boldsymbol x})$. Because ${\cal B}({\boldsymbol x})$ is real and symmetric it is diagonalizable by an orthogonal matrix ${\cal U}({\boldsymbol x})$ with ${\cal U}({\boldsymbol x}){\cal U}^T({\boldsymbol x})={\boldsymbol 1}$, where ${\boldsymbol 1}$ denotes the identity matrix. Moreover, the normalized eigenvectors ${\boldsymbol f}_i({\boldsymbol x})$ of ${\cal B}({\boldsymbol x})$ form an orthonormal basis, ${\boldsymbol f}_i({\boldsymbol x}) \cdot {\boldsymbol f}_j({\boldsymbol x}) = \delta_{ij}$. Thus, we can construct the transformation matrix ${\cal U}({\boldsymbol x})=({\boldsymbol f}_1({\boldsymbol x}), \ldots,{\boldsymbol f}_{d-1}({\boldsymbol x}))$ such that ${\cal B}({\boldsymbol x})={\cal U}({\boldsymbol x}) \Lambda({\boldsymbol x})  {\cal U}^T({\boldsymbol x})$ where $\Lambda({\boldsymbol x})$ is a diagonal matrix with the eigenvalues $\lambda_i({\boldsymbol x})$ of ${\cal B}({\boldsymbol x})$ along its diagonal. 
From \eq{diffmat} follows that ${\cal B}({\boldsymbol x})$ is positive definite and hence
 all eigenvalues $\lambda_i({\boldsymbol x})$ are positive. 
Here, we tacitly assume that all $T_{j k}$ are nonzero; if 
certain transitions are excluded,
 ${\cal B}({\boldsymbol x})$ is positive semidefinite and
 eigenvalues can be zero.

 Finally, our matrix ${\cal C}({\boldsymbol x})$ is given by  
\begin{equation}
{\cal C}({\boldsymbol x}) = {\cal U}({\boldsymbol x}) \cdot 
\left(
\begin{array}{ccc}
  \sqrt{\lambda_1} & \ldots  & 0   \\
 \vdots & \ddots  & \vdots   \\
 0 &  \ldots  &    \sqrt{\lambda_{d-1}}
\end{array}
\right)
 \cdot {\cal U}^T({\boldsymbol x}).
 \end{equation}
This standard procedure to diagonalize matrices can be easily implemented numerically. However,  the diffusion matrix ${\cal B}({\boldsymbol x})$ obviously depends on the transition probabilities and thus on the abundances of all strategies. Consequently, the procedure to calculate ${\cal C}({\boldsymbol x})$ must be continuously repeated as time progresses and the state $\boldsymbol x$ changes. This is computationally inconvenient and therefore, it is desirable to calculate  ${\cal C}({\boldsymbol x})$ analytically. 
 
The simplest case with $d=2$ strategic types results in a one-dimensional Fokker-Planck equation \cite{traulsen:2005hp}. Moreover, in special cases, for example in cyclic games such as the symmetric Rock-Scissors-Paper game, the Fokker-Planck equation~(\ref{eq:FPE}) can be approximated in polar coordinates \cite{reichenbach:2006aa}. 
However, such cases are non-generic and here we focus on general, higher dimensional situations with $d\geq3$. As a particular example to illustrate the framework numerically, we provide a detailed analysis of a generic Rock-Scissors-Paper game.

\section{Multiplicative noise for specific processes}
In general, the diffusion matrix ${\cal B}({\boldsymbol x})$ depends not only 
on the frequencies $\boldsymbol x$ but also on the payoffs (fitness) of the different strategic types. In this case, an analytic representation of ${\cal C}({\boldsymbol x})$ is only of limited use because it would be valid just for one particular game. However, ${\cal B}({\boldsymbol x})$ becomes payoff independent if $T_{j k}({\boldsymbol x}) + T_{k j}({\boldsymbol x})$ is payoff independent. 
Fortunately, this holds for the broad and relevant class of pairwise comparison processes. In these processes, a focal individual $f$ and a model $m$
with payoffs $\pi_f$ and $\pi_m$ are picked at random and a payoff comparison determines whether the focal individual switches its strategy. 

Let $\gamma(\pi_f,\pi_m)$ be the probability that the
focal individual adopts the strategy of the model 
(see e.g.\ \cite{sandholm:2010bo,wu:2010aa}) 
and assume that every
mutation leads to a \emph{different} strategy.
Then the transition probabilities from type $k$ to type $j$ read
\begin{equation}
T_{kj}({\boldsymbol x}) = (1-\mu)  x_k x_j \gamma(\pi_j,\pi_k) + 
\mu x_k
\frac{1}{d-1} 
\end{equation}
for $j \neq k$. 
Consequently, any pairwise comparison process with 
\begin{equation}
\label{eq:noisereq}
\gamma(\pi_j,\pi_k)+\gamma(\pi_k,\pi_j) = {\rm const.}
\end{equation}
leads to a payoff independent diffusion matrix $ {\cal B}({\boldsymbol x})$. 
If \eq{noisereq} is fulfilled, the noise term in \eq{langevin} is independent
of the evolutionary game. In particular, it allows
the consideration of multi-player games in
which the payoff functions are non-linear \cite{gokhale:2010pn}.
Examples for evolutionary processes that fulfill \eq{noisereq} include the local update process \cite{traulsen:2005hp} with $\gamma(\pi_j,\pi_k)=\frac12 + w(\pi_j-\pi_k)$ where $w$ indicates the strength of selection acting on payoff differences between different strategic types. For $w\ll1$ selection is weak, payoff differences amount to little changes in fitness and the process is dominated by random updating. For larger $w$ selection strength increases but 
an upper limit is imposed on $w$ by the requirement
 $0 \leq \gamma(\pi_j,\pi_k) \leq 1$. Another example is the Fermi process \cite{blume:1993jf,szabo:1998wv,traulsen:2006bb} with $\gamma(\pi_j,\pi_k) = \left( 1+\exp[-w(\pi_j-\pi_k)] \right)^{-1}$. Again $w$ indicates the selection strength but without an upper bound. In the limit $w\to\infty$, i.e. $\gamma(\pi_j,\pi_k) = \Theta\left[ \pi_j-\pi_k \right]$ where $\Theta[x]$ denotes the Heavyside step function, the imitation dynamics \cite{schlag:1998aa,traulsen:2009aa} is recovered. A further example, where 
$\gamma(\pi_j,\pi_k)$ does not simply depend on the difference between its arguments, 
is $\gamma(\pi_j,\pi_k) = \frac{\pi_j}{\pi_j+\pi_k}$ \cite{hauert:2002mn,nowak:2004pw}. 
Incidentally, all examples above satisfy $\gamma(\pi_j,\pi_k)+\gamma(\pi_k,\pi_j) = 1$ but, for example, there might be resilience to change in the local update process such that $\gamma(\pi_j,\pi_k)=\alpha\left(\frac12 + w(\pi_j-\pi_k)\right)$ with $0<\alpha<1$ and hence $\gamma(\pi_j,\pi_k)+\gamma(\pi_k,\pi_j) = \alpha<1$. 

Last but not least, an example of an important process that does not lead to a payoff independent ${\cal B}({\boldsymbol x})$, is given by the standard frequency dependent Moran process \cite{nowak:2004pw,traulsen:2008aa,altrock:2009aa} or its linearized equivalent \cite{claussen:2007aa,bladon:2010ws}.

In the following, we concentrate on cases with $\gamma(\pi_j,\pi_k)+\gamma(\pi_k,\pi_j) = 1$. This leads to
\begin{equation}
{\cal B}_{jk}({\boldsymbol x}) \!\!
= \!\! \frac{1}{N} \!\!
\left[
-x_j x_k(1-\mu)-\frac{\mu}{d-1}(x_j+x_k) 
 \right] \quad {\rm for} \; j\neq k
\end{equation}
and
\begin{equation}
{\cal B}_{jj}({\boldsymbol x}) \!\!
= \!\! \frac{1}{N} \!\!
\left[
x_j(1\!-x_j)(1\!-\mu)+
\frac{\mu}{d-1}(1+x_j (d-2)) 
 \right]\!. 
\end{equation} 
Unfortunately, even in this case, a full derivation of ${\cal C}({\boldsymbol x})$ is difficult for general $d$.
Thus, we focus on the more manageable but highly illustrative case of $d=3$. 
Henceforth, we set $x_1=x$ and $x_2=y$ (and $x_3=1-x-y$) for convenience.
\begin{figure}[thbp]
\includegraphics[angle=0,width=\linewidth]{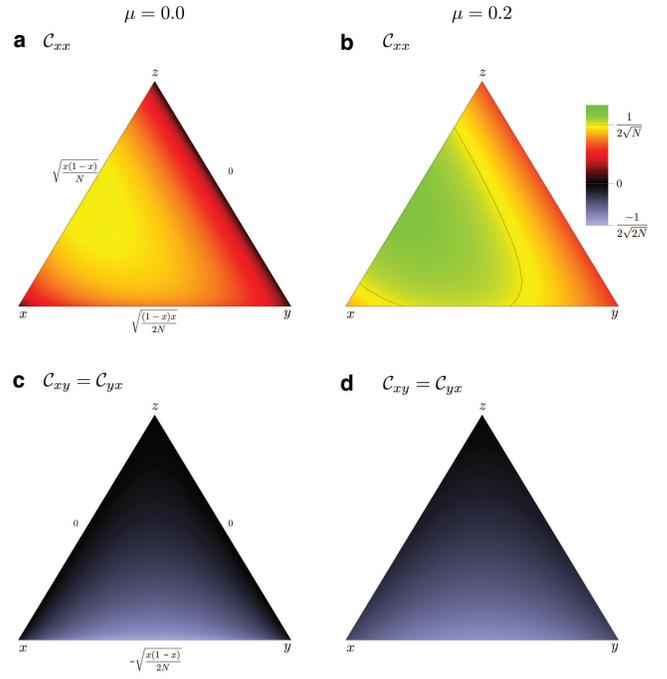}
\caption{
(Color online)  
Elements of the noise matrix ${\cal C}(x,y,z)$ for processes fulfilling \eq{noisereq} with $d=3$ strategies for $\mu=0$ (left) and $\mu>0$ (right). 
\textbf{a}
The element ${\cal C}_{xx}$ determines how the noise in the $x$-direction affects the
$x$-coordinate. 
In the case of $\mu=0$, this noise vanishes for $x \to 0$. 
For $y \to 0$ and $z\to0$ we recover the usual multiplicative noise from one-dimensional evolutionary processes 
\cite{traulsen:2005hp}.
\textbf{b}
For nonvanishing mutations ${\cal C}_{xx}$ increases
at $x \to 0$ and $x \to 1$ and exceeds the value 
$\tfrac{1}{2\sqrt{N}}$ (contour line)
in the interior of the simplex, which is the maximum value without mutations.
The element ${\cal C}_{yy}$ follows from the transformation $x \leftrightarrow y$. 
\textbf{c}
The element ${\cal C}_{xy}$ determines how the noise in the $x$-direction affects the
$y$-coordinate (or vice versa). 
This term is always negative, as explained in the text.
For $x \to 0$ or $y \to 0$, ${\cal C}_{xy}$ vanishes in the case of $\mu=0$. 
For $z \to 0$, we find ${\cal C}_{xy}=-{\cal C}_{xx}$, which ensures that the sum of the noise in the $x$-coordinate, ${\cal C}_{xx}+{\cal C}_{xy}$, vanishes on this edge of the simplex:
If the noise increases $x$ for $z=0$ it has to decrease $y$ by the same amount. 
\textbf{d}
In the case of $\mu>0$,  the noise no longer vanishes at the boundaries of the simplex
but the expressions are too cumbersome to display explicitly.
}
\label{fig:noise01}
\end{figure}

\subsection{No mutations, $\boldsymbol{\mu=0}$}
For $d=3$ and in the absence of mutations, we can give a relatively compact analytic expression for ${\cal C}({\boldsymbol x})$,
\begin{equation}
{\cal B}({\boldsymbol x}) = \frac{1}{N}
\begin{pmatrix}
 x (1-x) & -x y		\\
 -x y & y(1-y)
\end{pmatrix}.
\end{equation}
The eigenvalues of this matrix are
\begin{align}
\lambda_{\pm}({\boldsymbol x}) = \frac{K_+ \pm L}{2 N },
\end{align}
where 
$K_{\pm} = x(1-x) \pm y(1-y)$
and
$L = \sqrt{K_-^2 + 4 x^2y^2}$.
The eigenvectors are given by
\begin{align}
{\boldsymbol f}_{\pm}({\boldsymbol x}) = \frac{1}{\cal N} \left( -K_{-} \pm L, 2 x y \right),
\end{align}
where ${\cal N} = \sqrt{ 4x^2y^2+\left(  -K_{-} \pm L \right)^2}$ is a normalization factor. 
From this, it is straightforward to construct the transformation matrix ${\cal U}({\boldsymbol x})=({\boldsymbol f}_+({\boldsymbol x}),{\boldsymbol f}_-({\boldsymbol x}))$. 
The product ${\cal U}({\boldsymbol x}) \sqrt{\Lambda({\boldsymbol x})}\ {\cal U}^T({\boldsymbol x})$
then yields our matrix ${\cal C}({\boldsymbol x})$, which can be written as
\begin{subequations}
\begin{align}
{\cal C}_{xx}({\boldsymbol x}) & = 
\frac{1}{\sqrt{2 N}} 
\left( 
\frac{  \sqrt{Q_{+-}}  }{1+\left( \frac{2 x y}{ Q_{--} }\right)^2 }
+
\frac{ \sqrt{Q_{++}}  }{1+\left(\frac{2 x y}{ Q_{-+} }\right)^2 }
\right) 
\\
{\cal C}_{xy}({\boldsymbol x}) & =  {\cal C}_{yx}  = \frac{x y}{\sqrt{2 N } L}
\left( 
\sqrt{Q_{+-}} - \sqrt{ Q_{++}}
\right)
\\
{\cal C}_{yy}({\boldsymbol x}) & = 
\frac{1}{\sqrt{2 N}} 
\left( 
\frac{  \sqrt{Q_{+-}}  }{1+\left(\frac{2 x y}{ Q_{-+} }\right)^2 }
+
\frac{ \sqrt{Q_{++}}  }{1+\left(\frac{2 x y}{ Q_{--} }\right)^2 }
\right) 
\end{align}
\end{subequations}
where
$Q_{\pm \pm} = K_{\pm} \pm L$.
It is obvious that ${\cal C}_{xx}({\boldsymbol x}) \geq 0$ and ${\cal C}_{yy}({\boldsymbol x}) \geq 0$.
Since $L \geq 0$, we have $Q_{++} \geq Q_{+-}$ and therefore ${\cal C}_{xy}({\boldsymbol x}) \leq 0$. 
It is remarkable that even for this simple ${\cal B}({\boldsymbol x})$, the matrix ${\cal C}({\boldsymbol x})$ already takes a rather complicated form that is not easy to interpret. 
Therefore, the elements ${\cal C}_{xx}({\boldsymbol x})$ and ${\cal C}_{xy}({\boldsymbol x})$ are plotted in \fig{noise01}.
The remaining element, ${\cal C}_{yy}({\boldsymbol x})$, follows from the symmetry of the matrix,
i.e.\ in \fig{noise01} a,~b the simplex needs to be mirrored along the vertical axis $x=y$. 
Recall that the matrix ${\cal C}({\boldsymbol x})$ is independent of the evolutionary game that is played and 
does not depend on the microscopic evolutionary process, as long as 
\eq{noisereq}, $\gamma(\pi_j,\pi_k)+\gamma(\pi_k,\pi_j) = {\rm const.}$, holds.
While analytical calculations of ${\cal C}$ are also, in principle, feasible for $d=4$ and $d=5$, 
they lead to very lengthy expressions.

\subsection{With mutations, $\boldsymbol{\mu>0}$}
The procedure to derive ${\cal C}({\boldsymbol x})$ remains the same when including mutations just starting with
\begin{align}
{\cal B}({\boldsymbol x}) &= \frac{1\!-\mu}N
\begin{pmatrix}
 x (1\!-x) & -x y   \\
 -x y & y(1\!-y)
\end{pmatrix} \nonumber\\ &+
\frac\mu{2N}
\begin{pmatrix}
1\!+x & -x-y   \\
 -x-y & 1\!+y
\end{pmatrix}.
\end{align}
Unfortunately, the analytic expressions for the elements of ${\cal C}({\boldsymbol x})$ grow to unwieldy proportions. 
A graphical illustration of the elements of ${\cal C}({\boldsymbol x})$ for $\mu >0$ is shown in \fig{noise01}. 
Compared to $\mu=0$, the entries of ${\cal C}_{xx}({\boldsymbol x})$ become larger and 
the noise terms 
no longer vanish at the boundaries. 
For example, at the corners of the simplex we have for ${\cal C}_{xx}({\boldsymbol x})$
\begin{align*}
{\cal C}_{xx}(1,0,0) &= \tfrac{3}{\sqrt{10}}\sqrt{\tfrac\mu N} \\
{\cal C}_{xx}(0,1,0) &= \sqrt{\tfrac{2}{5}}\sqrt{\tfrac\mu N}\\
   {\cal C}_{xx}(0,0,1) &=\tfrac{1}{\sqrt{2}}       \sqrt{\tfrac\mu N }  
\end{align*}
For ${\cal C}_{xy}({\boldsymbol x})$, we obtain at these points
\begin{align*}
\nonumber
{\cal C}_{xy}(1,0,0) &= {\cal C}_{xy}(0,1,0)=
-\tfrac{1}{\sqrt{10}}
\sqrt{\tfrac\mu N }   \\ \nonumber
    {\cal C}_{xy}(0,0,1) &=0. 
\end{align*}
In addition, the functional form
of the noise term is altered with increasing $\mu$ and becomes significantly more complex than in the case of no mutations, $\mu=0$. 

For $\mu>0$ it is important to mention that for small mutation rates serious mathematical intricacies arise in the vicinity of absorbing boundaries and saddle-node fixed points \cite{chalub:2009tp,chalub:2009aa}. 
Intuitively, the reason for these complications 
arises from the diffusion approximation that underlies the derivation of the Fokker-Planck equation~(\ref{eq:FPE}). In systems with such absorbing 
boundaries (or sub-spaces where one or more strategic types are absent) finite populations spend non-negligible time on these boundaries in the limit $\mu\ll 1/N$. In contrast, the diffusion approximation is based on a continuum approximation, which reflects
the limit $N\to\infty$, and adds finite size corrections for large, finite $N$. For example, the state  of the population $\boldsymbol x$ is a continuous variable and hence can be located arbitrarily close to a boundary in both the Fokker-Planck or Langevin formalisms, \eq{FPE} and \eq{langevin}, but this is impossible in finite populations. 
More specifically, this implies that the case of $\mu\ll 1/N$, for which the population could get
trapped on an absorbing boundary for an extended period of time,
cannot be captured by the diffusion approximation. 
Consequently, the quality of the approximation is expected to decrease for small $\mu>0$ and 
to get worse if $N$ is small too.
Interestingly, this failure of the continuum approximation has not been raised in population genetics, 
where similar considerations have been made, but typically the description is only made on
the level of the Fokker-Planck equation and not based on stochastic differential equations 
\cite{ewens:2004qe}. However, population
geneticists are typically interested in simpler scenarios, where each type has a fixed fitness. 
In this case, the corresponding deterministic system has no generic trajectories that are close to 
absorbing boundaries. Moreover, the usual diffusion approximation in population genetics, where
the selection intensity scales to zero while the population size diverges, leads to substantial noise, such
that these effects have a minor impact.

For evolutionary games with small mutation rates one is typically not forced to apply this
relatively complex approximation. 
Instead, a time scale separation between mutation and selection
allows to approximate the dynamics based on a pairwise consideration of strategies
by considering an embedded Markov chain over the (quasi absorbing) homogenous states of the population
\cite{fudenberg:2006ee,imhof:2005oz,van-segbroeck:2009mi,sigmund:2010aa,wu:2011aa}.

\section{Application to the Rock-Paper-Scissors game}
To compare our approach based on stochastic differential equations in $d-1$ dimensions to individual based simulations with $d$ strategies, we focus on the case of $d=3$ where we have obtained closed analytical results above for $\mu=0$. 
As an example, we consider the cyclic dynamics in the Rock-Scissors-Paper-Game, 
which is not only a popular children's game, but also relevant in biological \cite{sinervo:1996le,sinervo:2006aa,czaran:2002ya,kerr:2002xg} and social systems \cite{hauert:2002te,semmann:2003he}. 
Moreover, the evolutionary dynamics of this game is theoretically very well understood, both in infinite as well as in finite populations \cite{hofbauer:1998mm,szabo:2002mf,reichenbach:2006aa,reichenbach:2007aa,claussen:2008aa,sandholm:2010bo}.
Here we focus on a generic Rock-Scissors-Paper game with payoff matrix
\begin{equation}
\label{eq:Mmatrix}
	\bordermatrix{
		  & R & S & P \cr
		R & 0 & \frac{s}{2} & -1 \cr
		S & -1 & 0 & 2+s \cr
		P & \frac{1+s}{3} & -1 & 0 \cr
		},
\end{equation}
which avoids artificial symmetries. According to the replicator equation, the game exhibits saddle node fixed points at $x=1$, $y=1$, and $z=1-x-y=1$ as well as an interior fixed point at $\hat{\boldsymbol x} = \left(\frac12,\frac13,\frac16\right)$, independent of the parameter $s$ 
but $s$ controls the stability of $\hat{\boldsymbol x}$. For $s>1$, $\hat{\boldsymbol x}$ is a stable focus and an unstable focus for $s<1$.
We do not consider the non-generic case of $s=1$, which exhibits closed orbits \cite{hofbauer:1998mm}. 

\begin{figure}[thbp]
\includegraphics[angle=0,width=0.6\linewidth]{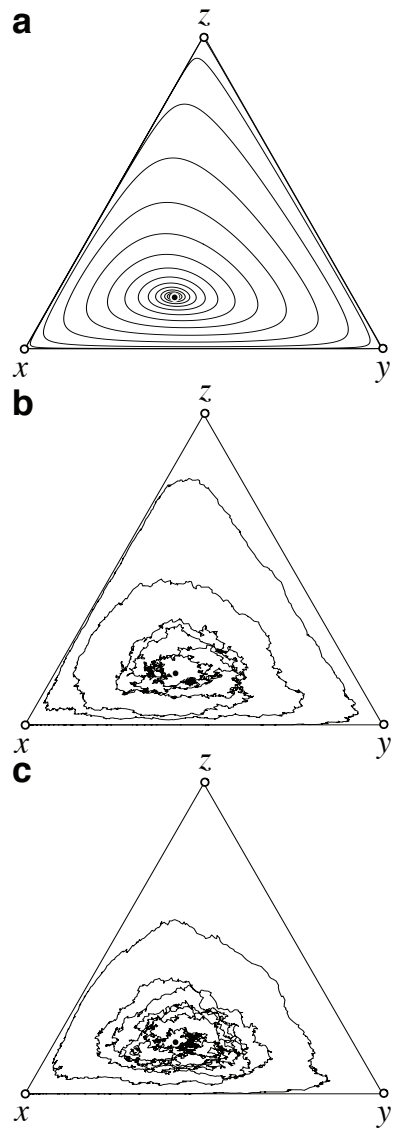}
\caption{
Comparison of trajectories for the {\bf a} deterministic case (replicator equation), {\bf b} stochastic differential equation, \eq{langevin}, and {\bf c} individual based simulations for an RSP interaction with weakly attracting interior fixed point $\hat {\boldsymbol x}$ with $s=1.4$ (c.f. \eq{Mmatrix}) and $N=1000$ (in {\bf b}, {\bf c}). Stochastic trajectories start close to $\hat {\boldsymbol x}$ and end at time $T=349.9$ (\textbf{b}) and $T=395.8$ (\textbf{c}).
Simulations can also be performed online at \cite{hauert:website:2012}.
}
\label{fig:traj}
\end{figure}

\begin{figure}[thbp]
\includegraphics[angle=0,width=\linewidth]{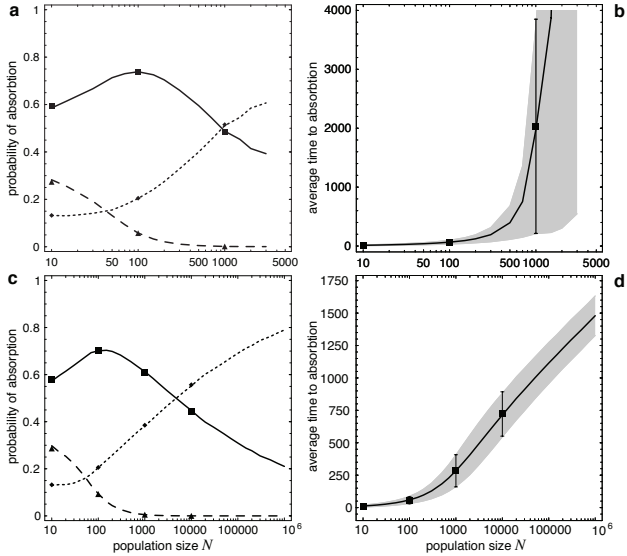}
\caption{
Probability of absorption and time to absorption (in the absence of mutations, $\mu=0$). \textbf{a}, \textbf{c} Probability to reach one of the three absorbing homogenous states of \emph{rock} (solid line, {\scriptsize$\blacksquare$}), \emph{scissors} (dashed line, $\blacktriangle$), or \emph{paper} (dotted line, {\scriptsize$\blacklozenge$}) as well as \textbf{b}, \textbf{d} the associated time to absorption for \emph{rock} as a function of the population size $N$ with a stable (top row, $s=1.4$) and unstable (bottom row, $s=0.8$) interior fixed point $\hat{\boldsymbol x}$. If $\hat{\boldsymbol x}$ is an attractor (\textbf{a}, \textbf{b}) it may take exceedingly long times for large $N$ until the population reaches an absorbing state -- even though this will inevitably occur. Because of this results are only shown up to $N=3000$. No such limitations occur if $\hat{\boldsymbol x}$ is a repellor. The symbols indicate results from individual based simulations. Error bars and grey shaded areas indicate the standard deviation of the mean for simulations and stochastic calculations, respectively.
Simulation results and results based on the stochastic differential equation \eq{langevin} were both averaged over $10^5$ independent runs.
}
\label{fig:nomut}
\end{figure}

Let us first analyze the fixation probabilities and the fixation times for the case without mutations $\mu=0$ for a game with $s=1.4$, such that $\hat{\boldsymbol x}$ is an attractor. 
The replicator equation, obtained in the limit $N \to \infty$, predicts that fixation never occurs and that the population evolves towards $\hat{\boldsymbol x}$, see \fig{traj}a.
However, in the stochastic system ultimately two strategies are lost, see \fig{traj}b,~c. Most importantly, the sample trajectories generated by the stochastic differential equation (\ref{eq:langevin}) in \fig{traj}b do not exhibit any qualitative differences when compared to individual based simulations, \fig{traj}c.
As a complementary scenario, we consider a game with $s=0.8$, such that $\hat{\boldsymbol x}$ is a repellor. According to the replicator equation trajectories now spiral away from $\hat{\boldsymbol x}$ towards the boundaries of the simplex and approach a heteroclinic cycle. The probabilities that the population reaches any one of the three homogenous absorbing states in either scenario is shown in \fig{nomut} together with the average time to fixation. 

\begin{figure}[thbp]
\includegraphics[angle=0,width=\linewidth]{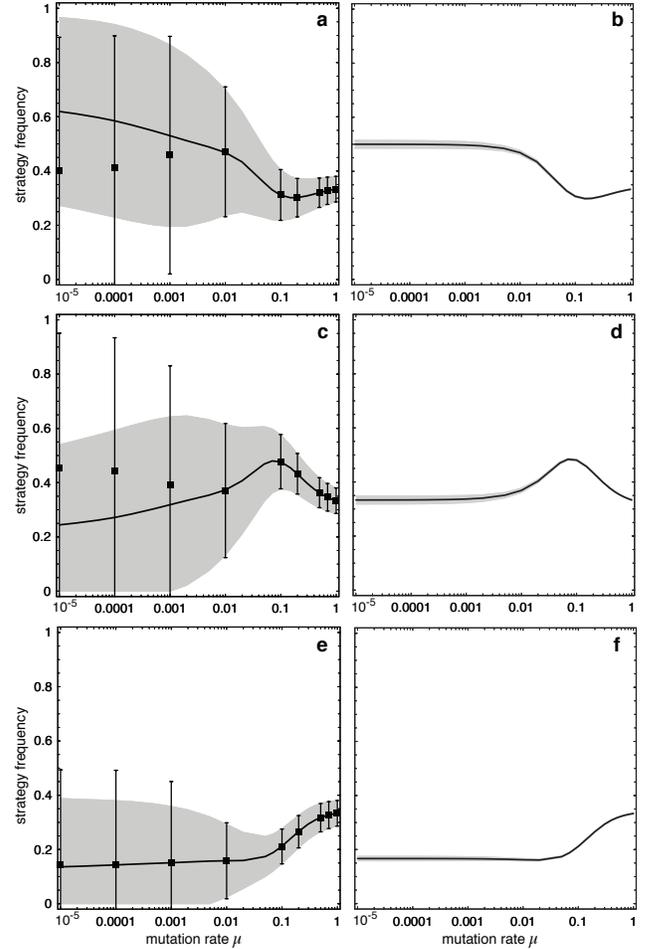}
\caption{
Average frequency (solid line) and standard deviation (grey shaded area) of the three strategies \emph{rock} (\textbf{a}, \textbf{b}), \emph{scissors} (\textbf{c}, \textbf{d}), and \emph{paper} (\textbf{e}, \textbf{f}) as a function of the mutation rate $\mu$ for small (left column, $N=100$) and large populations (right column, $N=10^5$) with a stable interior fixed point $\hat{\boldsymbol x}$, $s=1.4$, based on \eq{langevin}. For comparison, symbols and error bars depict results from individual based simulations. No simulation results are shown for $N=10^5$ because of prohibitive computational efforts. Simulations and \eq{langevin} are in excellent agreement for $\mu\gtrsim1/N$ but for smaller $\mu$ substantial deviations occur (see text for details). Stochastic fluctuations decrease with increasing $N$ and for $N=10^5$ the population spends most of the time in the close vicinity of $\hat{\boldsymbol x}$.
Simulation results and results based on the stochastic differential equation \eq{langevin} were both averaged over $10^{10}$ time steps after a relaxation time of $10^6$ steps ($dt=0.01$ for the Langevin equation).
}
\label{fig:mut}
\end{figure}
In all cases excellent agreement between the Langevin framework~(\ref{eq:langevin}) and individual based simulations is obtained. Note that larger $N$ not only improve the approximation \eq{langevin} but also result in a performance gain as compared to individual based simulations. More specifically, the computational effort scales linearly with population size for simulations but 
remains constant when integrating \eq{langevin} numerically.

For non-vanishing mutation rates, $\mu>0$, absorption is no longer possible and the boundary of the simplex becomes repelling. \fig{mut} depicts the average frequencies of each strategic type as a function of $\mu$. For small $\mu$ the population can still get trapped for considerable time along the boundary, which results in systematic deviations between the stochastic differential equation (\ref{eq:langevin}) and individual based simulations, \fig{mut}. More specifically, for $\mu<1/N$ the deviations increase with decreasing $\mu$ but the agreement remains excellent for $\mu>1/N$. This threshold simply means that for larger $\mu$ the time spent along the boundary can be neglected. In the limit $N\to\infty$ the replicator-mutator \eq{langevin} exhibits a stable limit cycle \cite{arenas:2011tb}.

\begin{figure}[thbp]
\includegraphics[angle=0,width=\linewidth]{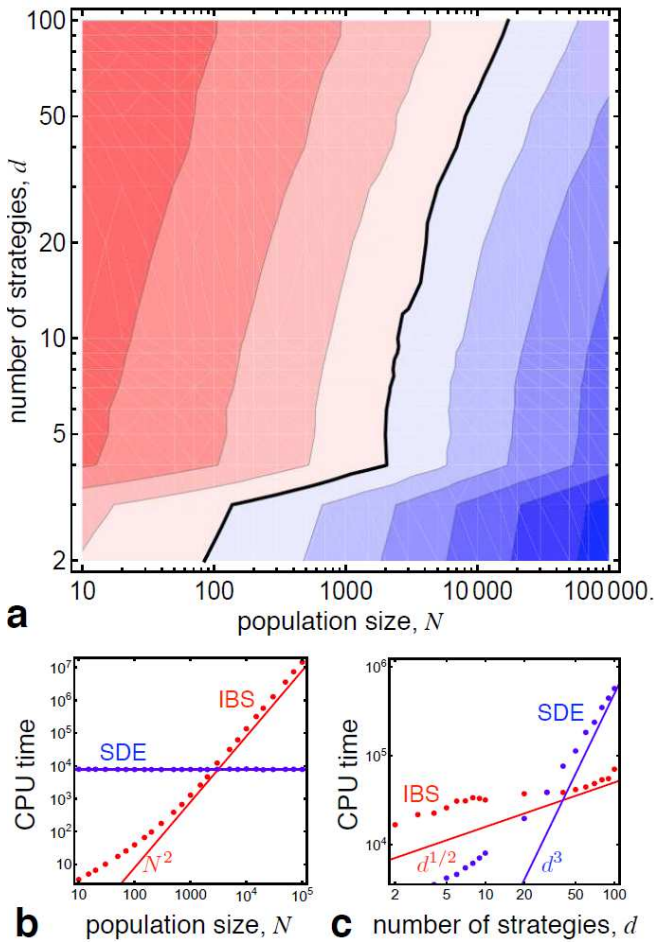}
\caption{(color online)
Performance comparison of individual based simulations (IBS) versus stochastic differential equations (SDE).
 \textbf{a} ratio of the CPU times $CPU_\text{SDE}/CPU_\text{IBS}$ as a function of the population size, $N$, and the number of strategic types, $d$. 
 The bold contour indicates equal performance. 
 For small $N$ and large $d$ IBS are faster (left of solid line), but for larger $N$ and smaller $d$ SDE are faster (right of solid line). 
 Each contour indicates a performance difference of one order of magnitude. 
 Note that IBS for $d=2, 3$ are based on analytical calculations of the eigenvalues/eigenvectors of ${\cal B}({\boldsymbol x})$, which requires substantially less time and hence explains the discontinuity in the contours. 
 \textbf{b} computational time with $d=10$ as a function of $N$ for IBS and SDE. 
 As a reference for the scaling $N^2$ and a constant are shown. \textbf{c} computational time with $N=5000$ as a function of $d$ for IBS and SDE. As a reference for the scaling $d^{1/2}$ and $d^3$ are shown. For a proper scaling argument much larger $d$ are required but already $d=100$ far exceeds typical evolutionary models and hence is only of limited relevance in the current context. All comparisons use a constant payoff matrix and the local update process \cite{traulsen:2005hp} 
(such that ${\cal A}_{k}({\boldsymbol x})=0$ and
 $\gamma(\pi_j, \pi_k)=1/2$), a mutation rate of $1/N$ and are based on at least $1000$ time steps as well as at least one minute running time. CPU time is measured in milliseconds required to calculate $1000$ time steps. The time increment for the SDE is $dt=0.01$.}
\label{fig:perf}
\end{figure}

\section{Discussion}
Evolutionary dynamics can be implemented in multiple ways. 
In particular, if only two types are present, the dynamics reduces
to a single dimension
and is typically solvable analytically, even if the
population is finite and demographic noise is present. 
If more than two types are present, it is significantly more challenging to 
describe the dynamics analytically. 
When the mutation rates are sufficiently small \cite{wu:2011aa}, there are typically at most two 
types present at 
any time, thus leading to situations which justify simpler tools 
based on the interaction of two types.
In the limit of infinite populations, deterministic differential equations arise and 
can be used to describe the system even when the population size is large, but finite. 
However, it is more challenging to incorporate noise in this case.
Here, we have proposed a way to address this issue by deriving a Langevin equation
for more than two types.
For large populations, it is typically much more efficient to solve these equations numerically than to resort to individual based simulations. A detailed performance comparison is shown in Fig.\ \ref{fig:perf}. Computational costs of simulations increase slowly with the number of strategic types, $\sim d^{1/2}$, but increase with the population size as $N^2$. In contrast, SDE's are essentially unaffected by $N$, but computational costs arise from repeatedly solving for eigenvalues and eigenvectors of ${\cal B}({\boldsymbol x})$. In theory, analytical solutions are available for $d\leq5$ but are probably meaningful in practice only for $d=2,3$ and numerical methods are required for $d>3$, which scale with $d^3$ \cite{demmel:1997bo}. To illustrate this, for the data point $N=10000$ in \fig{nomut}c,~d the simulations required approximately two months ($1418$ hours) to complete as compared to $49$ minutes for the corresponding calculation based on stochastic differential equations, which corresponds to a $1700$-fold performance gain.
However, if only a small number of one type is present in a population our approach does not work well unless there are no mutations or the product of the population size and the mutation rate is sufficiently high, $N \mu >1$, such that the time spent along the boundary becomes negligible, cf.\ \fig{mut}. 

In summary, our approach 
establishes a transparent link between deterministic models and individual-based simulations of evolutionary processes. Moreover, the resulting stochastic differential equations provide substantial
speed-up compared to simulations, and therefore may serve well
in the investigation of multidimensional evolutionary dynamics.

\acknowledgments{
We thank P.M.\ Altrock and B.\ Werner for helpful comments. 
A.T.\ acknowledges support by the Emmy-Noether program of the DFG and by the MPG. 
C.H.\ acknowledges support by the Natural Sciences and Engineering Research Council of Canada (NSERC).
}

\clearpage


\end{document}